\documentclass[twocolumn,letter]{jpsj2}

\def\runauthor{M. Matsumoto and M. Koga}

\title{Exciton Mediated Superconductivity in PrOs$_4$Sb$_{12}$}

\author{Masashige Matsumoto and Mikito Koga$^1$}
\inst{Department of Physics, Faculty of Science, Shizuoka University, 836 Oya, Shizuoka 422--8529, Japan \\
$^1$Department of Physics, Faculty of Education, Shizuoka University, 836 Oya, Shizuoka 422--8529, Japan}

\recdate{February 19, 2004}

\abst{
The most important character of the exotic superconductor PrOs$_4$Sb$_{12}$
is the existence of low-lying excitations (excitons) with a finite energy gap
and it appears as the magnetic field-induced order above 4.5 T.
We focus on the $a_u$ conduction band, which hybridizes with a Pr $4f^2$ state strongly,
coupled to the excitons.
It results in an attractive interaction between the conduction electrons.
The symmetry of the superconducting order parameter is determined
by dispersion relation of the exciton.
For the bcc system PrOs$_4$Sb$_{12}$,
a $d$-wave state
($k_x k_y + \omega k_y k_z + \omega^2 k_z k_x$, $\omega={\rm e}^{\pm {\rm i} 2\pi/3}$)
is stabilized with broken time reversal symmetry.
}

\kword{
skutterudite,
PrOs$_4$Sb$_{12}$,
superconductivity,
time reversal symmetry,
crystal field,
exciton,
heavy fermion
}

\begin{document}
\sloppy
\maketitle

\renewcommand{\theequation}{\arabic{equation}} 

\newcommand{\br}{{\mbox{\boldmath$r$}}}
\newcommand{\bk}{{\mbox{\boldmath$k$}}}
\newcommand{\sk}{{\mbox{\footnotesize $k$}}}
\newcommand{\bskp}{{\mbox{\footnotesize \boldmath$k$}}}
\newcommand{\bsrp}{{\mbox{\footnotesize \boldmath$r$}}}
\newcommand{\bsk}{\bskp}
\newcommand{\bsr}{\bsrp}
\newcommand{\bS}{{\mbox{\boldmath$S$}}}
\newcommand{\bs}{{\mbox{\boldmath$s$}}}
\newcommand{\bQ}{{\mbox{\boldmath$Q$}}}
\newcommand{\bsigma}{{\mbox{\boldmath$\sigma$}}}
\newcommand{\Pra}{PrOs$_4$Sb$_{12}$}
\newcommand{\PrRu}{PrRu$_4$Sb$_{12}$}
\newcommand{\La}{LaOs$_4$Sb$_{12}$}
\newcommand{\Uran}{UPd$_2$Al$_3$}

\Pra~is a recently found superconductor in the Praseodymium-based heavy-fermion system.
\cite{Bauer}
Specific heat measurement revealed multiple superconducting transition temperatures
at $T_{{\rm c1}}=1.85$ K and $T_{{\rm c2}}=1.75$ K.
\cite{Vollmer}
The experiment on thermal conductivity also reported the multiple phases
depending on temperature and external magnetic field.
\cite{Izawa}
The nuclear quadrupole resonance (NQR) experiment showed
that there is no Hebel-Slichter peak in $T_1^{-1}$
at the superconducting transition temperature.
\cite{Kotegawa}
Very recently, zero-field $\mu$SR measurement revealed
that the superconducting state is associated with a spontaneous magnetic field,
indicating that the superconducting state breaks the time reversal symmetry.
\cite{Aoki-muSR}
Thus far, only phenomenological theories have tried
to account for these experimental results of unconventional superconductivity.
\cite{Miyake,Goryo,Maki,Ichioka,Sergienko}
Since the multiple superconducting phase is still under investigation,
we do not discuss it in this letter.
The main purpose of this letter is to present a microscopic theory
considering electronic states specific to \Pra,
and to give a scenario for the superconductivity with broken time reversal symmetry.

The Pr$^{3+}$ ion has 4$f^2$ configuration in a $T_h$ point group crystal field.
\cite{Takegahara}
It is reported that the Fermi surface of \Pra~is similar
to that of \La~which is a reference compound,
\cite{Sugawara}
indicating well-localized 4$f^2$ nature.
This localized nature of the f-electrons is characteristic to \Pra,
compared with other heavy fermion superconductors with itinerant $f$-electron nature
such as in U-based compounds.
Therefore, we consider the conduction electron system well-separated from the $f$-electrons.

Another characteristic point of \Pra~is the magnetic field-induced ordered phase above 4.5 T,
observed by specific heat,
\cite{Aoki-field}
electric conductivity, magnetization and thermal expansion measurements.
\cite{Ho}
In this ordered phase,
high-field neutron scattering measurements revealed
a small staggered moment perpendicular to the field.
\cite{Kohgi}
Shiina and Aoki proposed that the field-induced order
is mainly driven by a quadrupole-quadrupole interaction.
\cite{Shiina-preprint}
They assumed the $\Gamma_1$ singlet ground state
and the $\Gamma_5$ triplet first excited state in Pr 4$f^2$ configuration
(specifically, $\Gamma_4$ should be used for $T_h$ representation).
Since an external magnetic field lifts the degenerate $\Gamma_5$ triplet
and stabilizes one of them,
this $\Gamma_1$-$\Gamma_5$ level scheme explains the field-induced order.
In this letter, we assume this $\Gamma_1$-$\Gamma_5$ scheme.

\PrRu~is a reference superconductor
($T_{\rm c}=1.04$ K \cite{Takeda})
with a Hebel-Slichter peak in $T_1^{-1}$ of NQR,
\cite{Yogi}
indicating an $s$-wave pairing state.
It is reported that \PrRu~has also triplet excitations as in \Pra.
\cite{Frederic}
However, a field-induced order has not been reported thus far in \PrRu.
This means that the crystal-field excitation gap to the triplet state
is much smaller in \Pra~than in \PrRu.
The low-energy excitations play important roles in the field-induced order
and the heavy electron mass for \Pra.
The low-lying excitation (exciton) is expected to be the most important origin
of the exotic superconductivity in \Pra,
while this is not the case for \PrRu.
In this letter, we present a microscopic theory for time reversal breaking superconductivity
mediated by excitons, specific to the bcc system \Pra.

The wave functions for the Pr 4$f^2$ state are
\begin{align}
|\Gamma_1 \rangle &= \frac{\sqrt{30}}{12} ( |4 \rangle + |-4 \rangle )
                   + \frac{\sqrt{21}}{6} |0 \rangle, \cr
|\Gamma_5^1 \rangle &= \sqrt{\frac{7}{8}} |3 \rangle - \frac{1}{\sqrt{8}} |-1 \rangle, \cr
|\Gamma_5^2 \rangle &= \frac{1}{\sqrt{2}} ( |2 \rangle - |-2 \rangle ), \cr
|\Gamma_5^3 \rangle &= -\sqrt{\frac{7}{8}} |-3 \rangle + \frac{1}{\sqrt{8}} |1 \rangle, \\
|\Gamma_4^1 \rangle &= -\frac{1}{\sqrt{8}} |-3 \rangle - \sqrt{\frac{7}{8}} |1 \rangle, \cr
|\Gamma_4^2 \rangle &= \frac{1}{\sqrt{2}} ( |4 \rangle - |-4 \rangle ), \cr
|\Gamma_4^3 \rangle &= \frac{1}{\sqrt{8}} |3 \rangle + \sqrt{\frac{7}{8}} |-1 \rangle.
\nonumber
\label{eqn:base}
\end{align}
Here, $\Gamma_1$, $\Gamma_4$ and $\Gamma_5$ are $O_h$ representations.
The total angular momentum is fixed to 4,
and the wave function $|J_z\rangle$ ($J_z=-4 \sim 4$)
represents the $z$ component of the angular momentum.
We take the basis functions $|\phi_n \rangle$ ($n=0,1,2,3$) for the $T_h$ system as
\cite{Shiina-preprint}
\begin{align}
&|\phi_0 \rangle = |\Gamma_1 \rangle, \cr
&|\phi_{n>0} \rangle = \sqrt{1-d^2} |\Gamma_5^n \rangle + d |\Gamma_4^n \rangle,
\end{align}
where $-1/\sqrt{2}\le d \le 1/\sqrt{2}$.
As an effective Hamiltonian for the 4$f^2$ states,
we take an intersite interaction into account as well.
\begin{align}
&H_f = H_{\rm CF} + H_{\rm I} \cr
&H_{\rm CF} = \Delta_{\rm CF} \sum_{n=1}^3 |\phi_n\rangle \langle\phi_n| \label{eqn:HPr} \\
&H_{\rm I} = \sum_{\alpha=1,2,3} \sum_{\langle ij \rangle} \sum_{\beta}
             D_{\beta \beta}  X^\alpha_{i\beta} X^\alpha_{j\beta}
\nonumber
\end{align}
Here, $H_{\rm CF}$ is a local term with a crystal field excitation gap $\Delta_{\rm CF}$.
We choose isotropic multipole-multipole interactions for $H_{\rm I}$ in order to
simplify the following discussion.
In $H_{\rm I}$, $D_{\beta \beta}$ is a coupling constant and
$X_{\beta}^\alpha$ is one of the three-dimensional tensor operators (denoted by $\alpha$)
for a dipole, a quadrupole and an octupole, among others (denoted by $\beta$).
$\sum_{\langle ij \rangle}$ denotes the summation
over the nearest-neighbor Pr sites in the bcc lattice.

First, we introduce bosonic operators such as
\cite{Kusunose,Shiina-2003}
\begin{equation}
X_\beta^\alpha = \sum_{n,n'=0}^3 x_{nn'\beta}^\alpha a_n^\dagger a_{n'},
\end{equation}
with $x_{nn'\beta}^\alpha = \langle\phi_n| X_\beta^\alpha |\phi_{n'}\rangle$
and $a_n^\dagger a_{n'} = |\phi_n\rangle \langle\phi_{n'}|$.
The bosons are subjected to $\sum_{n=0}^3 a_n^\dagger a_n =1$ at each site.
Due to the finite excitation gap,
bosons $a_{n>0}$ for excitations are dilute at low temperatures,
and we eliminate $a_0$ for the ground state using
$a_0 = a_0^\dagger = \sqrt{ 1 - \sum_{n=1}^3 a_n^\dagger a_n}$.
The multipole operator is rewritten up to the quadratic order in dilute boson operators as
\begin{align}
X_\beta^\alpha &= x_{00\beta}^\alpha + \sum_{n=1}^3
   ( x_{n0\beta}^\alpha a_n^\dagger + x_{0n\beta}^\alpha a_n ) \cr
&~+ \sum_{n,n'=1}^3 ( x_{nn'\beta}^\alpha - x_{00\beta}^\alpha \delta_{nn'} )
    a_n^\dagger a_{n'} + O(3).
\end{align}

One of the relevant intersite interactions in PrOs$_4$Sb$_{12}$ is of the $\Gamma_5$
quadrupolar type ($X_{\rm Q}^1=O_{xy}$, $X_{\rm Q}^2=O_{yz}$, $X_{\rm Q}^3=O_{zx}$).
\cite{Shiina-preprint}
\begin{equation}
O_{\xi \eta} = \frac{\sqrt{3}}{2} ( J_\xi J_\eta + J_\eta J_\xi )
\end{equation}
Using this quadrupole operator, we demonstrate a derivation of exciton dispersion.
Nonzero matrix elements for the quadrupole operators $x_{nn'\beta}^\alpha$
($\beta={\rm Q}$) are given as
$x_{20{\rm Q}}^1 = -{\rm i} x_{\rm Q}$, $x_{10{\rm Q}}^2 = {\rm i} x_{\rm Q}/\sqrt{2}$,
$x_{30{\rm Q}}^2 = -{\rm i} x_{\rm Q}/\sqrt{2}$,
$x_{10{\rm Q}}^3 = x_{\rm Q}/\sqrt{2}$, $x_{30{\rm Q}}^3 = x_{\rm Q}/\sqrt{2}$,
and $x_{\rm Q}=\sqrt{35(1-d^2)}$.
$H_f$ is now derived up to the quadratic order in the boson operators as
\begin{align}
&H_f = \sum_i \Delta_{\rm CF} \sum_{n=1}^3 a_{in}^\dagger a_{in} \\
&~+ \sum_{\langle ij \rangle}
\left[
  \lambda \sum_{n=1}^3 a_{in}^\dagger a_{jn}
+ \lambda'( a_{i1}^\dagger a_{j3}^\dagger + a_{i3}^\dagger a_{j1}^\dagger
       - a_{i2}^\dagger a_{j2}^\dagger ) + {\rm h.c.}
\right].
\nonumber
\end{align}
Here, $\lambda'=\lambda=x_{\rm Q}^2 D_{{\rm Q Q}}$ holds
for the only quadrupole-quadrupole interaction
in $H_{\rm I}$.
For a dipole-dipole interaction, $\lambda'=-\lambda$ is obtained.
In general, other multipoles are involved in the intersite interaction
as well as the quadrupole, so that $\lambda'$ differs from $\lambda$.
In our theory, the types of intersite interaction do not alter our result.
They just modify the constants $\lambda$ and $\lambda'$.

Since $a_1$ couples with $a_3$ by the pair creation and annihilation terms,
we transform the operators as
$a_1=-(a_x -{\rm i} a_y)/\sqrt{2}$,
$a_2=a_z$,
$a_3=(a_x +{\rm i} a_y)/\sqrt{2}$.
$H_f$ is then given by the decoupled bosons $a_\alpha$ ($\alpha=x,y,z$).
\begin{align}
&H_f = \sum_{\alpha=x,y,z} H_\alpha
\label{eqn:Hamiltonian} \\
&H_\alpha = \sum_{\bsk} [ e_{{\bsk}} a_{{\bsk}\alpha}^\dagger a_{{\bsk}\alpha}
        - \frac{1}{2} \Lambda_{\bsk } ( a_{{\bsk}\alpha}^\dagger a_{-{\bsk}\alpha}^\dagger + {\rm h.c.} ) ] \cr
&e_{{\bsk}}=\Delta_{\rm CF}+ \lambda \varepsilon_{{\bsk}},~~~~~~
\Lambda_{\bsk} = \lambda' \varepsilon_{{\bsk}} \cr
&\varepsilon_{{\bsk}}=8 \cos{\frac{k_x}{2}} \cos{\frac{k_y}{2}} \cos{\frac{k_z}{2}}
\nonumber
\end{align}
$a_{i\alpha}^\dagger = (1/\sqrt{N}) \sum_{\bsk} {\rm e}^{{\rm i} {\bsr}_i \cdot {\bsk}} a_{\bsk\alpha}^\dagger$
was introduced with $N$ as the number of Pr sites.
We can now diagonalize the Hamiltonian using a Bogoliubov transformation.
\begin{align}
&H_\alpha = \sum_{\bsk} E_{\bsk} b_{{\bsk}\alpha}^\dagger b_{{\bsk}\alpha}
          - \frac{1}{2} \sum_{\bsk} ( e_{\bsk} - E_{\bsk} ) \cr
&a_{{\bsk}\alpha} = u_{\bsk} b_{{\bsk}\alpha} + v_{\bsk} b_{-{\bsk}\alpha}^\dagger
\label{eqn:Bogoliubov} \\
&u_{\bsk} = \sqrt{\frac{1}{2}(\frac{e_{\bsk}}{E_{{\bsk}}}+1)},~~~
v_{\bsk} = {\rm sgn}(\Lambda_{\bsk}) \sqrt{\frac{1}{2}(\frac{e_{\bsk}}{E_{{\bsk}}}-1)}
\nonumber
\end{align}
Here, $b_{\bsk}$ describes low-energy bosonic excitations.
The dispersion relation of the exciton is given by
$E_{\bsk} = \sqrt{ e_{\bsk}^2 - \Lambda_{\bsk}^2}$.
There are threefold degenerate excitations ($\alpha=x,y,z$).
We note that the Hamiltonian (\ref{eqn:Hamiltonian}) is the same as that
for interacting spin dimer systems in which the field-induced order takes place.
\cite{Matsumoto}

Next, we derive an effective interaction between conduction electrons
via exciton creation and annihilation processes.
The conduction electron system has characteristics
of both the a$_u$ and t$_u$ molecular orbitals of the Sb$_{12}$ cage structure.
Since the a$_u$ component strongly couples with the 4$f^1$ wave function
of the 4$f^2$ $\Gamma_5$ triplet at the $\Gamma$ point,
\cite{Harima}
we restrict ourselves to the a$_u$ conduction band to discuss superconductivity.
We study the following Hamiltonian $H$.
\begin{align}
&H = H_c + H_f + H_{cf} \\
&H_c = \sum_{{\bsk}\sigma} \epsilon_{\bsk} c_{{\bsk}\sigma}^\dagger c_{{\bsk}\sigma},
~~~H_{cf} = -J_{cf} \sum_i \bs_{i}\cdot\bS_{i}
\nonumber
\end{align}
Here, $H_c$ is for the conduction electron system.
$H_f$ is given in eq. (\ref{eqn:Hamiltonian}).
$H_{cf}$ represents an effective exchange interaction ($J_{cf}>0$) between the conduction electrons
and the 4$f^2$ $\Gamma_5$ triplet.
\cite{Shiba}
The $\Gamma_1$ and $\Gamma_5$ states do not couple with each other via only spin exchange.
$H_{cf}$ is isotropic in $T_h$ symmetry.
$\bs_{i}$ denotes a spin ($S=1/2$) operator
for a conduction electron system at the $i$-th site.
\begin{equation}
\bs_i = \frac{1}{2} \sum_{\sigma\sigma'}
        c_{i \sigma}^\dagger \bsigma_{\sigma\sigma'} c_{i \sigma'}
\end{equation}
Here, $c_{i \sigma}$ is an annihilation operator for the a$_u$ electron at the $i$-th site.
$\bS_{i}$ is a pseudospin ($S=1$) operator for the $|\phi_{n>0}\rangle$ triplet,
which is given at each Pr site by
\begin{equation}
\bS = -{\rm i} (a_x^\dagger,a_y^\dagger,a_z^\dagger) \times (a_x,a_y,a_z).
\end{equation}
When conduction electrons excite the triplet excitons via $H_{\rm cf}$,
they induce polarization of the pseudospin $\bS$.
The other conduction electrons approach the polarized site in the next process,
giving rise to an effective interaction between the conduction electrons.
We treat $H_{cf}$ as a perturbation
and derive an effective interaction Hamiltonian up to the second order.
\begin{equation}
H' = H_{cf} \frac{1}{E_0 - H_c - H_f} H_{cf}
\end{equation}
Since the characteristic energy is $\Delta_{\rm CF} \sim 8$ K,
\cite{Maple,Kohgi}
the excitons are dilute at $T<T_{\rm c}=1.85$ K.
We consider only the exciton pair creation and annihilation process
which is dominant at low temperatures.
In this case, we obtain the following $H'$
using the Bogoliubov transformation (\ref{eqn:Bogoliubov}).
\begin{align}
&H' = \sum_{\bsk} V_{\bsk}~\bs_{\bsk}\cdot\bs_{-{\bsk}} \\
&V_{\bsk} = 2 J_{pf}^2 \frac{1}{N}
\sum_{{\bsk}'} \frac{u_{{\bsk}-{\bsk}'}v_{{\bsk}-{\bsk}'}u_{{\bsk}'}v_{{\bsk}'}- v_{{\bsk}-{\bsk}'}^2 u_{{\bsk}'}^2}
{ E_{{\bsk}-{\bsk}'} + E_{{\bsk}'} }
\nonumber
\end{align}
Here, $\bs_{\bsk}$ is the Fourier-transformed spin operator of the conduction electrons.
We neglected the energies of the conduction electrons in the denominator.
The excitation has a small dispersion,
\cite{Kuwahara}
namely $|\Lambda_{\bsk}| \ll \Delta_{\rm CF}$,
and $V_{\bsk}$ is simplified as
\begin{align}
&V_{\bsk} = V_0 ( \cos{\frac{k_x}{2}} \cos{\frac{k_y}{2}} \cos{\frac{k_z}{2}} -1 ),
\label{eqn:V0} \\
&V_0 = \frac{2 (\lambda'J_{cf})^2}{\Delta_{\rm CF}^3}.
\nonumber
\end{align}
In real space, $H'$ can be written as
\begin{equation}
H' = \frac{1}{4}V_0 \sum_{\langle ij \rangle} \bs_i \cdot \bs_j
- V_0\sum_i \bs_i \cdot \bs_i.
\end{equation}
The first term is an antiferromagnetic interaction
between the nearest-neighbor sites [along (111) directions],
while the second term is a ferromagnetic interaction on the same sites.
The former leads to a singlet pairing.
The latter is a ferromagnetic short-range (repulsive $s$-wave) interaction,
and does not contribute to triplet pairings.
We study then the effective Hamiltonian for the conduction electron system.
\begin{equation}
H_{\rm eff} = \sum_{{\bsk}\sigma} \epsilon_{\bsk} c_{{\bsk}\sigma}^\dagger c_{{\bsk}\sigma}
+ \sum_{\bsk} V_{\bsk}~\bs_{\bsk}\cdot\bs_{-{\bsk}}
\end{equation}

A simple mean-field analysis leads to the following gap equation
for singlet pairings at low temperatures.
\begin{equation}
\Delta(\bk) = \frac{1}{N}
\sum_{{\bsk}'} \frac{3}{2} V_{{\bsk}-{\bsk}'}^{({\rm e})}
\frac{\Delta(\bk')}{2\sqrt{\epsilon_{{\bsk}'}^2+|\Delta^2(\bk')|^2}}
\end{equation}
Here, $V_{{\bsk}-{\bsk}'}^{({\rm e})}$ represents the even component of $V_{{\bsk}-{\bsk}'}$
under $\bk\rightarrow -\bk$ or $\bk'\rightarrow -\bk'$ transformations.
It is expressed as
\begin{align}
&V_{{\bsk}-{\bsk}'}^{({\rm e})} = V_0 \{ [ f_s(\bk) f_s(\bk') -1 ] + f_{d1}(\bk) f_{d1}(\bk') \cr
&~~~~~~~~~~~~~~
                                 + f_{d2}(\bk) f_{d2}(\bk') + f_{d3}(\bk) f_{d3}(\bk')
                             \}, \\
&f_s(\bk) = \cos(k_x/2) \cos(k_y/2) \cos(k_z/2), \cr
&f_{d1}(\bk) = \sin(k_x/2) \sin(k_y/2) \cos(k_z/2), \cr
&f_{d2}(\bk) = \cos(k_x/2) \sin(k_y/2) \sin(k_z/2), \cr
&f_{d3}(\bk) = \sin(k_x/2) \cos(k_y/2) \sin(k_z/2).
\nonumber
\end{align}
For triplet pairings, we note that effective interactions are all repulsive.
For singlet pairings, there are $s$-wave [$f_s(\bk)f_s({\bsk}')-1]$
and $d$-wave [$f_{dn}(\bk)f_{dn}(\bk')]$ $(n=1,2,3$) channels.
The $d$-wave channel is attractive,
while the $s$-wave channel is repulsive due to the $-1$ term.

Now we determine what type of superconducting state is favorable for the $d$-wave.
For this purpose, we derive a Ginzburg-Landau free energy from the gap equation.
The gap equation for $d_1$, corresponding to $d_{xy}$-wave,
is written using the Matsubara frequency.
\begin{align}
&\eta_1 \propto \sum_{\omega_m} \int d \Omega_{{\bsk}'} f_{d1}(\bk')
  \frac{\Delta(\bk')}{\sqrt{\omega_m^2 + |\Delta(\bk')|^2}} \\
&\Delta(\bk) = \eta_1 f_{d1}(\bk) +  \eta_2 f_{d2}(\bk) +  \eta_3f_{d3}(\bk)
\nonumber
\end{align}
Here, $\eta_1$, $\eta_2$ and $\eta_3$ are complex numbers representing order parameters
for the $d_{xy}$, $d_{yz}$ and $d_{zx}$-wave, respectively.
$\int d\Omega_{{\bsk}'}$ means an integral over the Fermi surface.
We assume $|\Delta(\bk')|$ is small near $T_{\rm c}$, and expand the denominator.
The third-order terms are given by
\begin{align}
\frac{\delta F_4}{\delta \eta_1^*} &\propto
  \int d \Omega_{{\bsk}'} f_{d1}(\bk') |\Delta(\bk')|^2 \Delta(\bk') \\
&=
    A |\eta_1|^2 \eta_1
 +  2B ( |\eta_2|^2 + |\eta_3|^2 ) \eta_1
 +  B ( \eta_2^2 + \eta_3^2 ) \eta_1^* \cr
A &= \int d \Omega_{{\bsk}'} f_{d1}^4(\bk'),~~~
B = \int d \Omega_{{\bsk}'} f_{d1}^2(\bk') f_{d2}^2(\bk')
\nonumber
\end{align}
Here, $F_4$ is the fourth-order term in the free energy written as
\begin{align}
F_4 &\propto \frac{1}{2} A ( |\eta_1|^4 +|\eta_2|^4 +|\eta_3|^4 ) \cr
&~+ 2B (|\eta_1|^2 |\eta_2|^2 + |\eta_2|^2 |\eta_3|^2 + |\eta_3|^2 |\eta_1|^2 ) \\
&~+ \frac{1}{2} B ( {\eta_1^*}^2 \eta_2^2
                + {\eta_2^*}^2 \eta_3^2
                + {\eta_3^*}^2 \eta_1^2 + {\rm c.c.} ).
\nonumber
\end{align}
The generic form of $F_4$ for the three-dimensional representation is given by
\cite{Volovik}
\begin{align}
F_4 &= \beta_1 ( |\eta_1|^2 + |\eta_2|^2 + |\eta_3|^2 )^2
      + \beta_2 | \eta_1^2 + \eta_2^2 + \eta_3^2 |^2 \cr
&~+ \beta_3 ( |\eta_1|^4 + |\eta_2|^4 + |\eta_3|^4 ).
\end{align}
There are the following relations between the coefficients.
\begin{equation}
\beta_1 \propto B > 0,~~~~
\beta_2 \propto \frac{1}{2} B > 0,~~~~
\beta_3 \propto \frac{1}{2} (A-3B)
\end{equation}
For a simple spherical Fermi surface, $\beta_3 > 0$ holds.
In the most stable state,
the time reversal symmetry is broken with the following order parameter:
\cite{Volovik,Sigrist}
\begin{equation}
\Delta(\bk) = \eta [ f_{d1}(\bk) + \omega f_{d2}(\bk) + \omega^2 f_{d3}(\bk) ].
\label{eqn:Delta}
\end{equation}
Here, $\omega={\rm e}^{\pm {\rm i} 2\pi/3}$.
This state has eight point nodes.
There is eightfold degeneracy due to the four directions of the threefold axes
and to the time reversal degeneracy.

In this letter, we studied the exciton mediated superconductivity
to discuss the time reversal symmetry breaking state realized in \Pra.
Our idea is based on the fact that there exist low-lying excitations
above the nonmagnetic $\Gamma_1$ singlet ground state.
This is a unique feature of \Pra~among known skutterudite superconductors.
Our theory reveals that the superconducting pairing symmetry
is determined by the $\bk$ dependence of the exciton dispersion.
For a bcc lattice, a three-component $d$-wave state appears
with broken time reversal symmetry,
which agrees with the result of the zero-field $\mu$SR experiment.
We point out that the most stable state (\ref{eqn:Delta})
breaks the time reversal symmetry around one of the threefold axes
due to the relative phase ${\rm e}^{\pm {\rm i} 2\pi/3}$ between the three components.

A magnetic exciton mechanism was suggested as a origin of a $d$-wave superconductivity in \Uran.
\cite{Sato,Thalmeier,McHale}
In this case, the superconductivity coexists with antiferro-magnetic order.
In our model for \Pra, the ground state is a nonmagnetic $\Gamma_1$ singlet
which does not suffer superconductivity,
and we consider excitons with a finite energy gap.
These points are different from the model for \Uran.

Our theory can predict that the gapped-exciton-mediated superconductivity is suppressed
with the increase in crystal field excitation gap $\Delta_{\rm CF}$ [see eq. (\ref{eqn:V0})],
which is the characteristic energy for the excitons.
This can be realized by substituting Ru for Os,
since \PrRu~has a much higher crystal field excitation energy gap.
\cite{Frederic}
Very recently, it has been reported that $T_{\rm c}$ decreases from 1.85 K ($T_{\rm c}$ of \Pra)
with the substitution of Ru for Os,
and then it increases towards 1.04 K ($T_{\rm c}$ of \PrRu),
which is understood as a competition between the $d$-wave and $s$-wave.
\cite{Frederic}
The first decrease in $T_{\rm c}$ agrees with our theory.
Another way to increase the crystal field excitation gap $\Delta_{\rm CF}$
is by applying pressure.
In our model, the $T_{\rm c}$ of \Pra~should be decreased with pressure.

Throughout the letter, the conduction band is restricted to the $a_u$ band
hybridizing strongly with an $f$-electron wave function of the Pr 4$f^2$ state.
In fact, the $t_u$ band electron also couples with the 4$f^2$ state
and admixes with the $a_u$ on the Pr sites,
which leads to an orbital exchange interaction.
This interaction couples the Pr $\Gamma_1$ singlet ground state to the $\Gamma_5$ triplet,
and the orbital exchange (or interband exchange) process is involved
in the effective interaction between conduction electrons.
The possibility of multiband superconductivity will be studied in the future.

The idea of gapped-exciton-mediated superconductivity can be adopted to spin dimer systems,
where a nonmagnetic singlet ground state is realized
accompanied by dispersive triplet excitations with a finite energy gap.
It is essentially the same as the Pr 4$f^2$ system discussed in this letter.
Introducing a conduction electron system coupled to the spin dimers,
we can expect superconductivity mediated by the triplet excitons.
This is one of the directions to search for a new superconductor.

We are indebted to Y. Aoki, K. Kuwahara, H. Shiba and R. Shiina for valuable discussions.


\end{document}